\documentclass[journal]{IEEEtran}
\ifCLASSINFOpdf
\else
\fi
\usepackage[utf8]{inputenc}
\usepackage{graphicx}
\usepackage{textcase}
\usepackage{blindtext}
\usepackage[numbers]{natbib}
\usepackage{array}
\usepackage{colortbl}
\usepackage{xcolor}

\begin{document}
%
\title{Multi-Tier Computing-Enabled Digital Twin in 6G Networks}
%
%
%

\author{Kunlun~Wang,~\IEEEmembership{Member,~IEEE,}
        Yongyi~Tang, 
        Trung~Q.~Duong,~\IEEEmembership{Fellow,~IEEE,}
        Saeed~R.~Khosravirad,~\IEEEmembership{Member,~IEEE,}
        Octavia~A.~Dobre,~\IEEEmembership{Fellow,~IEEE,}
and~George~K.~Karagiannidis,~\IEEEmembership{Fellow,~IEEE}
\thanks{Kunlun Wang and Yongyi Tang are with the School of Communication and Electronic Engineering, East China Normal University, Shanghai 200241, China, and also with the Joint Innovation Laboratory of Digital GuangXi Smart Infrastructures, Guangxi Information Center, Nanning 530000, China (e-mail: klwang@cee.ecnu.edu.cn); 51255904070@stu.ecnu.edu.cn).}
\thanks{Trung Q. Duong is with the School of Electronics, Electrical Engineering and Computer Science, Queen’s University Belfast, BT7 1NN Belfast, U.K. (e-mail: trung.q.duong@qub.ac.uk).}
\thanks{Saeed R. Khosravirad is with the Nokia Bell Laboratories, Murray Hill, NJ 07964 USA (e-mail: saeed.khosravirad@nokia-bell-labs.com).}
\thanks{Octavia A. Dobre is with the Faculty of Engineering and Applied Science, Memorial University of Newfoundland, St. John’s, NL A1C 5S7, Canada (e-mail: odobre@mun.ca).}
\thanks{George K. Karagiannidis is with the Department of Electrical and Computer Engineering, Aristotle University of Thessaloniki, 54 124 Thessaloniki, Greece and also with the Cyber Security Systems and Applied AI Research Center, Lebanese American University (LAU) (e-mail: geokarag@auth.gr).}
}

\maketitle
\begin{abstract}
Digital twin (DT) is the recurrent and common feature in discussions about future technologies, bringing together advanced communication, computation, and artificial intelligence, to name a few. In the context of Industry 4.0, industries such as manufacturing, automotive, and healthcare are rapidly adopting DT-based development. The main challenges to date have been the high demands on communication and computing resources, as well as privacy and security concerns, arising from the large volumes of data exchanges. To achieve low latency and high security services in the emerging DT, multi-tier computing has been proposed by combining edge/fog computing and cloud computing. Specifically, low latency data transmission, efficient resource allocation, and validated security strategies of multi-tier computing systems are used to solve the operational problems of the DT system. In this paper, we introduce the architecture and applications of DT using examples from manufacturing, the Internet-of-Vehicles and healthcare. At the same time, the architecture and technology of multi-tier computing systems are studied to support DT. This paper will provide valuable reference and guidance for the theory, algorithms, and applications in collaborative  multi-tier computing and DT.
\end{abstract}


%
\IEEEpeerreviewmaketitle

\section{Introduction}
%
%
%
%
With the deployment and advancement of fifth-generation wireless network (5G), academia and industry have begun to explore the relevant technologies, requirements and use cases of sixth-generation wireless network (6G), and digital twin (DT) is one of the key use cases. One of the key promises of the 6G networks is to connect the physical and digital worlds, which has brought DT to the forefront of research and development \cite{9923927}, \cite{viswanathan2020communications}. DT refers to the technology and methods for simulating and analyzing physical entities or systems based on digital models and real-time data. DT can transform real world entities or systems into digital forms, that enable real-time monitoring, prediction, and resource optimization in a virtual environment. These twin models include not only physical characteristics and structures, but also their associated processes, environments, and interactions. The aim of DT is to provide a highly reliable platform for product design and optimization, process monitoring, operational orchestration, and maintenance. In practice, DT has been  implemented in many industries, such as industrial automation, transportation, architectural design, and healthcare \cite{8740963, 10121437}. In order to realize the wide application of DT, it is necessary to meet the requirements of ultra-low latency, ultra-high reliability, and high privacy for data transmission between physical entities and DT entities \cite{9120192, 10183802, 9885226}. Traditional cloud computing models cannot fully meet these requirements, as massive data and computing tasks need to be transferred to the cloud through a forward transmission link, resulting in eavesdropping on private data. In addition, traditional cloud computing systems typically store data and compute tasks in the cloud, which greatly increases the latency of data processing. To provide low latency services and improve data security, a new computing model, namely multi-tier computing is proposed, which efficiently combines edge computing and cloud computing \cite{yang2018meets}, \cite{chiang2016fog}. 
\par In the 6G era, multi-tier computing systems can take advantage of the different characteristics of servers through proper multi-level coordination to support computing applications. In addition, a large number of intelligent devices with different computing resources around users can realize communication and computing resource sharing to perform task computation \cite{zhou2022digital}. In a multi-tier computing system, the cloud is the central node for overall management, providing large-scale computing and storage resources to support the processing of complex applications and algorithms. Edge computing, on the other hand, can serve as distributed nodes, providing computing and storage resources close to end devices, to support real-time data processing and response \cite{9606720}. Multi-tier computing is integrated with communication functions in 5G and beyond systems to meet the requirements of DT applications, including ultra-low latency, ultra-high reliability, energy efficiency and privacy protection. 
\par DT applications include a wide range of applications such as intelligent loT \cite{yang2019multi}, ad-hoc networks \cite{10078095}, intelligent manufacturing \cite{nain2022towards}, smart grid \cite{dragivcevic2019future}, dynamic network slicing \cite{10089865}, Internet-of-Vehicles (IoV)  \cite{9985947}, network slice \cite{9998964}, open radio access network \cite{10179151}, and smart cities \cite{anthony2020big}. These applications of DT require high-precision modeling of physical entities to accurately predict their behavior, and the continuous acquisition of data from physical entities for model updates and predictions. 
DT applications require high data transmission rates, for remote and real-time computation of tasks. However, end-users typically have limited computing, communication, and storage resources. To solve the problem of limited resources, multi-tier computing systems can offload computational tasks from users to different helper nodes with sufficient and unused resources. Task offloading in multi-tier computing systems allows distributed intelligent nodes to share their idle computing and communication resources, so that multi-dimensional resources can be efficiently used for low-latency task computation and computational efficiency can be improved \cite{9783171, 9786719}. In addition, DT contains sensitive entity information such as detailed characteristics of physical systems, work status, and historical data. Therefore, corresponding measures must be taken to protect the privacy and security of data \cite{9927259}. Multi-tier computing can implement measures such as data encryption \cite{9723011}, identity authentication \cite{8875291}, and access control \cite{faraji2014identity} at different levels of the system to ensure the data security. The development of 5G and future 6G wireless communication systems, as well as the new generation of embedded artificial intelligence (AI), will support DT through multi-tier computing systems. As computing capability and storage data shift from the cloud to edge nodes, computing and networking will be deeply integrated with the development of wireless communication systems. Therefore, the computing capability and security performance from the cloud to things should be better coordinated, which will promote the application of DT to a new level.
\subsection{Main Contributions}
This paper presents a vision for multi-tier computing enabled DT, focusing on its interactions with various wireless techniques and performance enhancements. Future research directions and open issues are discussed, covering DT application scenarios based on multi-tier computing.
\par In this context, our contributions are as follows:
\par $\bullet$ We elaborate on the definition and architecture of DT and provide a detailed introduction to its applications in manufacturing, automotive networking, and healthcare.
\par $\bullet$ We provide a detailed list of relevant problems and existing solutions in the next generation of wireless network-based multi-tier computing systems. 
\par $\bullet$ The research directions and future prospects of next generation wireless network enabled multi-tier computing systems empowered DT are discussed.

\subsection{Paper Organization}
This paper is organized as follows. Section \uppercase\expandafter{\romannumeral2} describes the application direction of DT and the challenges it faces, Section \uppercase\expandafter{\romannumeral3} proposes the empowerment of DT by multi-tier computing architecture. Section \uppercase\expandafter{\romannumeral4} mainly discusses the future directions and open problems of the combination of multi-tier computing and DT. The conclusions are drawn in Section \uppercase\expandafter{\romannumeral5}.
\section{Digital twin}
\subsection{Preliminaries of Digital Twin}
Digital twin was first introduced by Professor Grieves of the University of Michigan in 2003 speech on product life cycle management, which was the prototype of DT, and later published in a reference white paper, promoting the development of DT \cite{grieves2014digital}. In \cite{grieves2017digital}, DT is defined as a set of virtual information structures that comprehensively describe complex products from the micro to the macro world, to describe potential production or actual manufacturing of products. Any information that can be obtained by inspecting actually manufactured products can also be obtained from their DT. It is explicitly stated in \cite{grieves2017digital} that the purpose of DT is to use simulation to predict and eliminate the complexity of products and systems as possible, in order to avoid unpredictable and unacceptable behavior that could cause unknown disasters. Subsequently, the National Aeronautics and Space Administration defined DT as a process for aircraft or system development, that makes full use of the best physical models, sensors, and operational history data, integrates multidisciplinary and multiscale probabilistic simulation processes, and represents the corresponding physical state of the aircraft, marking a critical milestone for DT \cite{tuegel2011reengineering}. 
\par
\begin{figure*}[htp]
    \centering
    \includegraphics[width=14.5cm]{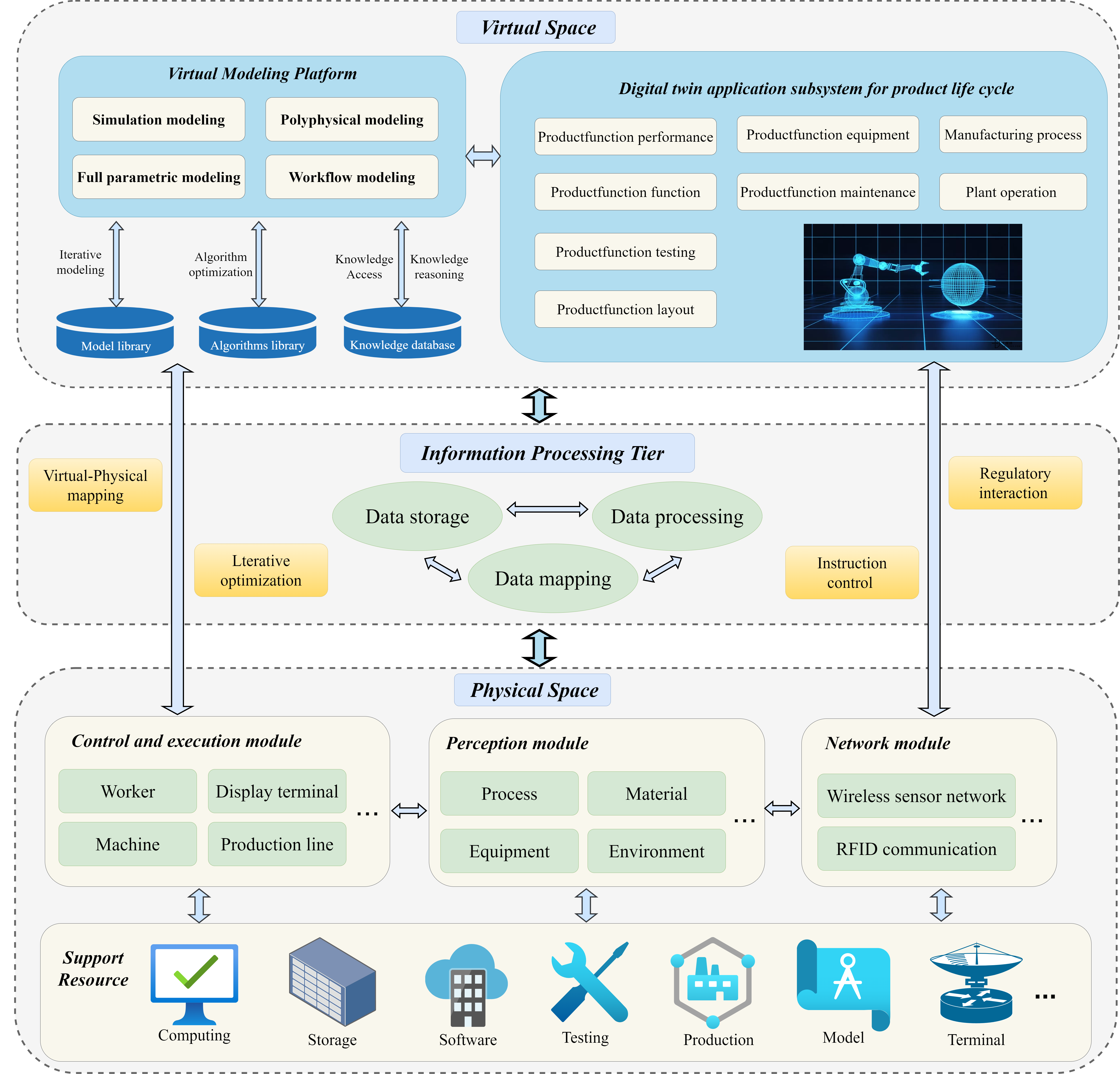}
    \caption{The structure of digital twin.}
    \label{555}
\end{figure*}

 The concept of DT requires at least four elements: a digital model, associated data, identification, and real-time monitoring capabilities. DT can be defined as a (physical and/or virtual) machine or computer-based model that simulates, emulates, mirrors or ``pairs" the life of a physical entity, which can be an object, a process, a person or a feature associated with a person. Each DT system is identified and linked to its physical entity by a unique specific key, thus allowing a bijective relationship to be established between the DT model and the physical entity \cite{kiritsis2011closed}, \cite{rios2015product}. A DT model is not just a simple model or simulation \cite{grieves2014digital}, \cite{kritzinger2018digital}, \cite{boschert2016digital},  \cite{negri2017review}. It is a living, intelligent and continuously evolving model that is the virtual counterpart of a physical entity or its process. It follows the life cycle of its physical twin to monitor, control and optimize its processes and functions. It continuously predicts future states (such as defects, damage, failures) and allows the simulation and testing of novel configurations for preventive maintenance. Fig. \ref{555} shows the DT structure. Table \ref{tab:1} summarizes the basic differences between simulation and digital twin.
\begin{table*}[htbp]
	\centering
	\caption{Comparison of Simulation and Digital Twin}
	\label{tab:1}  
	\begin{tabular}{m{3cm} m{5cm} m{5cm}}
		\hline\hline\noalign{\smallskip}	
		Feature & Simulation & Digital Twin  \\
		\noalign{\smallskip}\hline\noalign{\smallskip}
		\rowcolor{lightgray}Real-time Updates &  Real-time, accelerated, or decelerated. & Real-time updates, continuously receiving data from the corresponding real-world entity.  \\
		Purpose & Study complex system behavior, test hypotheses and parameters, evaluate performance and outcomes. & Monitor, diagnose, optimize, and predict the operation and performance of physical entities.  \\
		\rowcolor{lightgray}Application Areas & Aerospace, automotive engineering, climate simulation, financial risk assessment, etc. & Industrial manufacturing, urban planning, energy management, transportation, etc.  \\
		Data Source & User-specified input data or model. & Sensors, data collection, and internet technologies.  \\
		\rowcolor{lightgray}Associated Entities & Not necessarily connected to real-world entities. & Connected with real-world physical entities or systems.  \\
		Realism and Accuracy & Depends on the accuracy of the simulation model and input data. & Highly realistic and accurate, based on actual data and models.  \\
		\rowcolor{lightgray}Resource Requirements & May require significant computational resources and time. & Requires real-time data transmission and processing equipment, consumes higher resources.  \\
		Decision Support & Provides simulation results, supports system analysis and optimization. & Provides real-time data feedback, supports real-time decision-making.  \\
		\noalign{\smallskip}\hline
	\end{tabular}
\end{table*}
\par With today's greatly improved wireless communication capabilities, real-time data upload and big data storage capabilities enable the exchange and updating of descriptive data. The parameters of physical entities and surrounding environments are updated in real-time, and DT quickly obtains updated information from the real physical world. Through the application of data fusion algorithms, followed by big data analytics and artificial intelligence descriptive algorithms, it evolves with its physical model due to the modular and highly parameterized architecture, allowing rapid reconfiguration. Then, the DT continuously synchronizes and changes with its physical entity, and the changes reflect the attributes of the mirrored physical entity and are controlled by it. In addition to this comprehensive simulation, when combined with artificial intelligence, the DT can also discover information including system descriptions, hidden patterns, and unknown correlations. The ability to record, control, and monitor the state and changes of physical systems allows the application of AI predictive and prescriptive technologies to predict failures, test the results of potential solutions, and activate self-repair mechanisms. This leads to predictive maintenance methods, where failures can be predicted, and repair results can be simulated to avoid errors and find the best solution. The emergence of DT has greatly promoted the application of the Internet-of-Things (IoT) in the physical industry.

\subsection{Digital Twin Applications}
In this section, we introduce the DT application scenarios. With the progress of research on 5G and next-generation wireless network systems, the terminology and concepts of DT are continuously expanding in the academic community. With the advancement of IoT and next generation wireless networks, DT is being applied in various fields, such as education and training, digital governance, energy and utilities, aerospace and defence. The main application areas we focus on are the manufacturing industry, IoV, and medical systems.
\subsubsection{Manufacturing industry}
The main reason why DT is widely used in manufacturing is that manufacturers are always looking for ways to track and monitor their products in order to save costs. 
Additionally, the convenient connection between virtual models and physical entities is also one of the main motivations for the use of DT in the manufacturing industry. It can provide real-time status of physical equipment performance and production line feedback as much as possible, making it easy for managers to predict system behavior through feedback and adjust production methods in a timely manner. The use of DT will increase the connectivity and feedback between devices, thereby improving reliability and performance. The combination of AI algorithms and DT will improve the accuracy of model predictions, as prediction equipment sensors capture large amounts of real-time data, needed for performance and predictive analysis. 
\par This will greatly improve the accuracy of predictions, in contrast to traditional manufacturing environments where this information is usually stored in local databases on individual devices and only used to determine the cause of equipment failure after it has occurred \cite{9491874}. Currently, DT can be used to "support production system management" by building DT models called "basic management systems for managers", which help to structure and manage machine data \cite{kunath2018integrating}. For example, in \cite{liu2019digital}, a DT model is proposed  for identifying the design of an automated assembly line manufacturing system, which predicts the decision support information collected from dynamically executed intelligent multi-objective optimizations, and provides effective feedback for engineering solutions. At the same time, the real-time nature of DT can help "monitor and improve the production process." In \cite{wang2021digital}, an economical DT framework proposed, which consists of traditional machine structures but with tolerable industrial IoT capabilities, and used to monitor traditional machines in real-time, connect isolated machines to interconnected systems, and monitor machine status in real-time. DT technology can help manufacturing companies to optimize the entire lifecycle from design, manufacturing, maintenance, upkeep, and retirement, thereby improving production efficiency, quality, and sustainability. In \cite{konstantinov2017cyber}, a cyber physical systems is proposed for the entire lifecycle of a manufacturing system and product, which can be rapidly modified by a DT model to avoid bottlenecks in the development lifecycle caused by simulation, testing, and modification processes. Recycling and remanufacturing in the product lifecycle, face major challenges such as individual diversification, lack of product knowledge, and dispersed locations. DT provides personalized service systems, such as the novel DT system based on waste electrical and electronic equipment recycling proposed in \cite{wang2019digital}, which has been validated and evaluated in the implementation process in cloud and network physical systems. 
DT can be applied to the entire production process in the manufacturing industry, thus greatly promoting the development of the industry.
\subsubsection{Internet-of-Vehicles}
IoV can be seen as the integration of vehicular ad-hoc networks and the IoT, which connects various devices in vehicles (such as embedded processors and onboard units) to the smart road infrastructure in cities or highways \cite{tsukada2020networked}. These devices can sense and collect data (such as environmental and traffic) before sharing it with other devices (such as vehicles, roadside units, fog and edge devices, and cloud servers.) \cite{tsukada2020networked}, \cite{premsankar2018efficient, 10146005}. The data can then be processed and integrated with other data (such as user-generated data and open-source intelligence) to provide information for different levels of decision-making \cite{schwarting2018planning}. There are two possible directions for the application of DT in IoV, i.e., equipping individual vehicles with DT and equipping the traffic system with DT.
\par Configuring a DT for an individual vehicle enables real-time monitoring, diagnosis, and prediction of vehicle operation, which can optimize vehicle performance, extend lifetime, and improve reliability and safety. For example, \cite{Magargle2017ASD}, considers a model-driven approach to support thermal monitoring and predictive maintenance of automotive braking systems. This method involves creating a simulation-based DT or numerical model that combines different types of modeling into an integrated model of the brake system, that can be used for monitoring, diagnosis, and prediction. An integrated vehicle health management architecture for holistic vehicle health management is investigated in \cite{8914244}, which uses vehicle sensor inputs and diagnostic tasks to assess and predict the health status of critical components. In this paradigm, DT provides vehicle manufacturers with a greater ability to diagnose abnormal conditions and predict the remaining life of vehicle components without the need for on-site testing, thereby improving safety and customer satisfaction. DT can also be applied to vehicle maintenance and iteration. In \cite{bhatti2021towards}, the authors have discussed how DT can be used to optimize maintenance and repair, and how an intelligent interface can be built to reveal usage-based vehicle health degradation, supporting condition-based maintenance. From the perspective of product lifecycle management, vehicle DT enables virtual models to penetrate the entire vehicle lifecycle, including simulation, design, production, operation and disposal \cite{glaessgen2012digital}. In addition to the simulation of the vehicle itself, vehicle DT also includes the simulation of the driver. The vehicle collects the driver's behavioral habits through driving records, such as the driver's specific reaction in a given situation, thereby building a driver's behavior model and helping the entire in-vehicle network make smarter decisions in emergency situations. However, in order to predict anomalies such as driver fatigue or discomfort, the driver's mobile device must be connected to the in-vehicle network, which constantly updates the driver's real-time data (such as schedules and health information) through integrated dedicated monitoring applications \cite{liu2019edge}. 
\par Building a DT for a vehicle enables an efficient autonomous vehicle ecosystem. However, we still need to accurately simulate the external environment to maintain the safety and stability of the entire transport system through DT. In transport systems, DT can use real-time data and models to simulate traffic flow and predict future traffic conditions. It can also monitor the performance and status of transport systems and achieve rapid diagnosis and mitigation of traffic congestion \cite{9666030}. For example, in \cite{9860491}, a DT road network is proposed to observe the traffic operation from a macro perspective with the structure of convolutional long short term memory network (Conv-LSTM) units. The authors then stacked several Conv-LSTM layers to form a codec structure, which predicts the space-time congestion caused by accidents in the urban road network.

\subsubsection{Healthcare}
The main purpose of applying DT to healthcare is to construct human DT models that show what is happening inside the associated human. Then combined with the current environmental data (e.g. location, time, activity), disease prediction is made by analyzing the individual history of human DT. \cite{erol2020digital}. This will fundamentally change the paradigm of medical treatment, moving from treatment based on ``historical cases" to personalized treatment based on current personal ``body data", defined by all the structural, physical, biological, and historical health features of the individual. This has led to an emerging approach to disease treatment and prevention, the branch of healthcare that promotes personalized treatment, often referred to as "precision medicine" (more generally, "personalized medicine"). It involves the use of new diagnostic and treatment methods based on a patient's own recessive genetic disease, biomarker, phenotype, or psychological characteristics. Essentially, patients are viewed as individuals rather than according to some ``norms" or ``care standards" (more widespread pathological classification measures) \cite{schork2019artificial}. A specific implementation form is to monitor the patient's precise health condition in real-time through their health data, e.g., a model-based systems engineering approach is used in \cite{10027072} to create precise classification and accurate real-time simulation of health data assets using networked physical systems. By constructing a DT model of patient health data, hospitals can predict a patient's disease. Progression disease and formulate appropriate medical intervention, thus significantly improving the recovery rate of patients and the safety of treatment decisions.

\begin{figure}[htp]
    \centering
    \includegraphics[width=8.5cm]{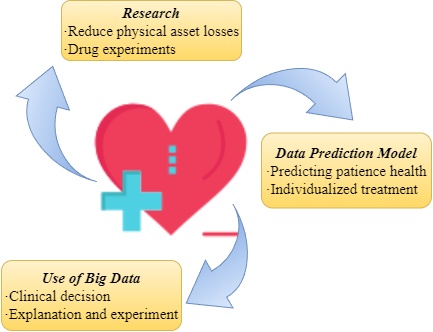}
    \caption{Application of DT technology for various aspects of healthcare industry.}
    \label{777}
\end{figure}

\par DT can not only directly monitor a patient's comprehensive health data, but also serve as a platform for researchers and clinicians to test all viable treatment options. With the widespread application of AI in medical scenarios, the continuous advancement of simulation and virtual reality technology, and the popularization of image archiving and communication systems, these provide economical storage and convenient access and exchange for the realization of multiple modalities (source machine type) medical examinations (mainly images). Nowadays, DT models of some organs (such as the heart) or parts of the human body (such as the respiratory system) have been developed \cite{martinez2019cardio}, \cite{bao2023patent}. Virtual physiological models will allow doctors to make computer predictions of all possible manifestations of real organs in any given situation. The automatic analysis provided by computer aided design systems will allow for the effectiveness of customized treatments to be evaluated, paving the way for the expansion of precise personal medicine \cite{sekhar2021brain}. DT technology can also be use prospectively to analyze the optimal pattern for modelling drugs, especially for individual patients \cite{rahman2022explore}. For example, in \cite{bahrami2022predicting}, a physical based DT for fentanyl transdermal therapy is proposed to find the appropriate dosage of medication based on the patient's pain feedback prediction.

\par In summary, DT technology has brought a number of benefits to healthcare, such as reduced clinical trial time, accurate preventive medical treatment plans, easier drug discovery processes, and fewer errors. Similarly, doctors will be able to provide better diagnosis and less invasive treatment. Fig. \ref{777} illustrates the application of DT in the healthcare industry
\subsection{Digital Twins' Challenging Questions}
As shown in the previous section, DT technology has been used in several applications. However, in order to make DT completely indistinguishable from physical twin, many challenges need to be overcome. In this section, we will introduce the challenges facing DT in terms of real-time two-way links between DTs and physical entities, data privacy and security issues, and computational resources.
\subsubsection{Delay}
In DT applications, real-time performance is a critical issue, especially in application scenarios involving control and scheduling, where latency can lead to system performance degradation or failure. The delay requirements of DT include two main aspects: transmission delay and computation delay. Transmission delay refers to the time required to collect, store, and transmit DT data \cite{mashaly2021connecting}.
\par Real-time data communication is one of the main factors affecting the quality of user experience in real-time communication. Due to the need for real-time data transmission between digital and physical twins to maintain their constant synchronization, the high latency of this communication process will result in inaccurate DT model or DT that cannot represent system updates in real-time. Firstly, the network topology and bandwidth limitations of data transmission may lead to an increase in transmission latency. Secondly, delays may be introduced during the collection and processing of sensor data, further affecting the real-time performance of DT. Therefore, several technologies that can be implemented to reduce communication latency:
\par $\bullet$ Deploying high-speed network links: using high-speed cables (such as fiber optic) and high-speed wireless network technology for wireless connections (such as 5G) is the first step in ensuring high-quality and high transmission rates. This reduces the transmission duration and waiting time \cite{chukhno2022placement}· 
\par $\bullet$ Data compression: Compressing data to reduce the amount of data to be sent is another way to reduce transmission latency. This can be achieved by using various data compression techniques. Alternatively, by deploying machine learning techniques that can detect changes in the system, only send them and omit any redundant data transmission without updates \cite{he2018surveillance}.
\par Computation delay refers to the time required to complete a task in a DT system, including data processing, algorithm calculation, control commands, etc. For DT, the problem of computational delay is essentially a problem of computing resource scheduling, which is introduced in the following section.
\subsubsection{Computing resources}
DT has certain computing resource requirements, especially when dealing with large-scale and complex data, performing real-time analysis, and simulations. The operation of DT involves the collecting, storing, and processing of massive real-time data, which requires efficient computing capability and storage systems to provide real-time responses. Additionally, simulating physical entities, predicting behaviors, and executing optimization algorithms in DT also demands substantial computational capability. It is impractical to simply rely on traditional stand-alone local servers for task computation, and it is necessary to integrate high-performance computing, distributed computing, cloud computing, and edge computing to build DT systems with sufficient computational resource \cite{sun2020reducing}, \cite{wang2022mobility}, \cite{wu2021digital, 10051857, 10077734}.
\par For example, cloud computing and edge computing can be employed. Cloud computing is a large-scale computing approach that leverages the Internet to enable on-demand sharing of computing and storage resources anytime, anywhere. DT that prioritizes computing speed and centralized processing can be deployed on cloud servers. Through cloud servers, large volumes of data can be processed in a short period, providing robust DT services. Furthermore, the cloud architecture facilitates the organization and management of numerous interconnected physical objects, virtual twin, real-time data, and historical experiences. In a cloud architecture, various types of storage devices can collaborate through application software to provide data storage and business access for enterprises \cite{tao2017digital}, \cite{qi2018modeling}. 
\par Edge computing, as a new paradigm of distributed computing, is recognized as a promising solution for addressing the limitations of computational resources in DT \cite{liu2022edge}. It involves the analysis and processing of specific data using computational, storage, and communication resources distributed along the path from the data source to the cloud center. However, storage and computing resources at edge nodes are limited. Data from virtual models require analysis by intelligent algorithms, which depend on adequate storage resources. The challenge lies in how to store different models at edge nodes for data analysis to meet the requirements of different physical objects. Efficiency in data analysis is crucial to ensure accurate feedback on physical objects. However, data analysis consumes significant computational resources, while edge nodes have limited computing resources. Computational offloading strategies can alleviate the pressure on edge nodes and greatly enhance the quality of data feedback on physical objects. Computational offloading refers to the transfer of computational tasks from end devices to other computing nodes with abundant computing resources to improve efficiency and reduce the burden of end devices. 
\par End devices often have limited computational resources and cannot handle complex computational requirements. By offloading computational tasks to computing nodes with abundant computing resources such as cloud servers, edge nodes, or dedicated computing platforms, the computational capabilities of these nodes can be fully utilized, resulting in improved computing performance and application response time \cite{9174795}. Computational offloading involves task allocation and scheduling considerations, taking into account factors such as the availability of computing resources, load balancing, and the cost of data transmission. By appropriately assigning computational tasks to appropriate computing resources, more efficient computing and lower computational latency can be achieved \cite{9676649}. However, it is still a challenge to design a task offloading strategy for DT and address the scarcity of computational resources \cite{9978919}.
\subsubsection{Privacy and security}
\begin{figure}[htp]
    \centering
    \includegraphics[width=8cm]{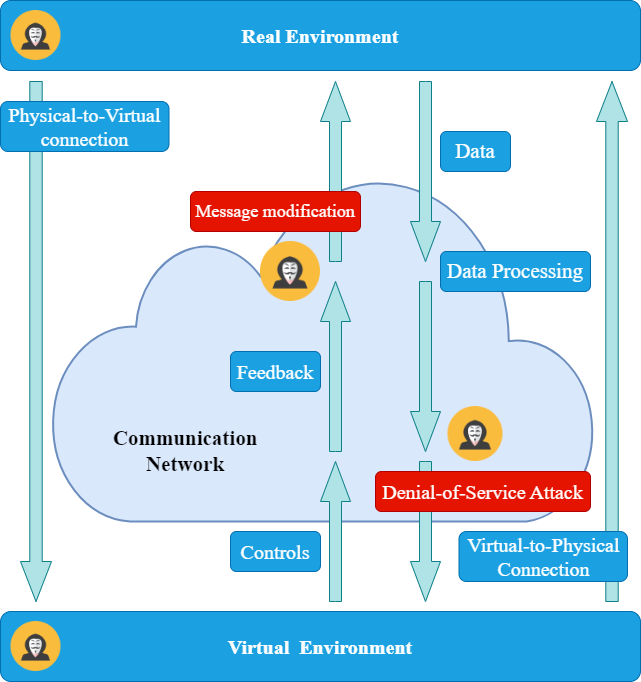}
    \caption{Links with security threats in DT.}
    \label{444}
\end{figure}

Data communication applications play a vital role in various components of DT technology. DT relies on technologies such as IoT and wireless communication technologies to obtain data from physical assets through sensors and various intelligent devices. The data in the DT needs to be collected, transmitted, and stored by relying on standard communication protocols, and the universal communication techniques used have well-known security issues \cite{qi2021enabling}. Therefore, the weakest link in DT security lies in data communication, while varies types of security threats exist in the overall environment. As shown in the Fig. \ref{444}, we provide the links where DT pose security threats.

\begin{figure*}[htp]
    \centering
    \includegraphics[width=12cm]{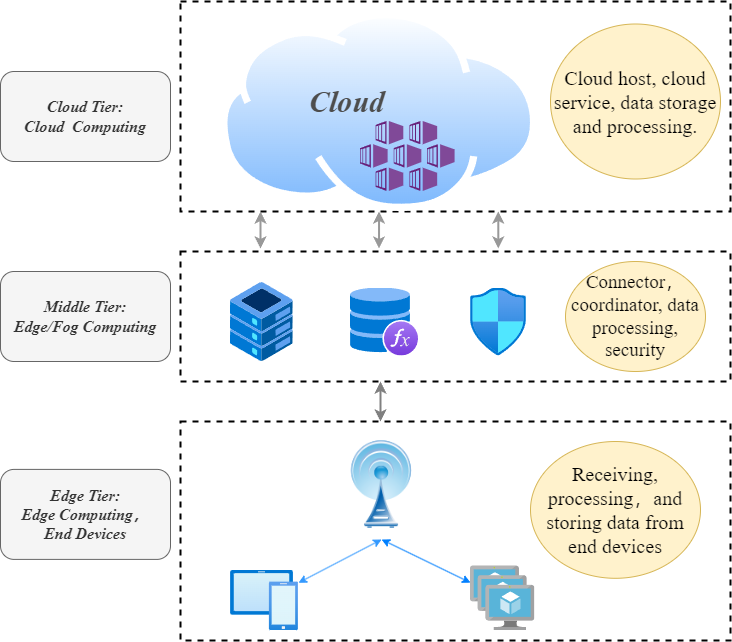}
    \caption{The structure of multi-tier computing system.}
    \label{666}
\end{figure*}

\par First, during the data collection phase, DT may face false data injection from sensors and node capture \cite{ahmed2020false}, \cite{wang2020understanding}. A false data injection attack is a method by which an attacker injects false data into a network or system by manipulating or forging data, in order to disrupt the normal operation of the system or obtain sensitive information. In a medical DT system, this attack changes patient DT disease information, increases drug demand, and makes the infusion pump deliver more medication than is prescribed \cite{jimenez2020health}. Node capture attack is one of the common attacks in wireless sensor networks, because nodes in wireless sensor networks are usually limited by resources, it is difficult to deal with such attacks. Therefore, the node capture attack is an important security issue in wireless sensor networks. In this case, the key node will be the microcontroller, as it aggregates and processes data to promote the generation of DT. In a patient monitoring system based on DTs, this attack method will lead to the theft of patient confidential data, so that the attacker can obtain the generation method of patient DT \cite{kaiser20206g}. In the face of such attacks, encryption mechanisms are often used to protect data from attackers. Examples of encryption algorithms are advanced encryption standard, data encryption standard, and Rivest-Shamir-Adleman method. However, unlike traditional communication models, DT has specific requirements that need to be considered, such as large data volumes, strict real-time constraints, and high data integrity \cite{wang2023design}.

\par Some of the potential security attacks that may occur during the DT data transmission phase are mainly man in the middle (MitM), sniffing, and denial-of-service (DoS) \cite{eckhart2019digital}. MitM attack refers to an attacker inserting themselves as a communication intermediary, impersonating a legitimate participant, intercepting, tampering, or monitoring communication traffic in a DT system. Attackers can modify data transmitted through intermediaries, thereby altering the state or results of the DT system, leading to incorrect decisions or operations. At the same time, sensitive information can also be leaked, such as identity credentials, digital assets, etc. \cite{mallik2019man}. Sniffing is a passive method where attackers can use sniffing tools such as networkminer to capture and analyze network packets to obtain information about network activity. However, in order to carry out this attack, the attacker must first obtain access privileges (physical connection or remote network access). Then there are location, network topology and time window issues, where the attackers need to be located at the right locations and time windows in the network to intercept the target data packets \cite{gluavan2020sniffing}. A DoS attack is a serious security threat to both the DT and the physical entity. Attackers send a large amount of task data information so that the DT system cannot respond normally \cite{donkal2021digital}. For example, in intelligent transportation, by launching DoS attack on the DT communication tier, attackers can shut down the network established by roadside units (RSUs) and prevent communication between vehicles and RSUs \cite{hussaini2022taxonomy}.
\par Lastly, there is the issue of data storage in DT, where most data storage operations occur in cloud computing. The stored data is sensitive and may face various risks, including the potential for data leaks if the cloud computing environment is targeted by attackers. Resolving access control issues is crucial to ensure that only authorized users or system components can access and modify DT data. Adequate access control policies and mechanisms, such as authentication and permission management, should be implemented to prevent unauthorized access \cite{putz2021ethertwin}.

\section{Multi-tier computing Empowered digital twin}
In this section, we first introduce the concept of multi-tier computing systems, and then propose that multi-tier computing systems empower DT to address latency, computing resource scheduling, and privacy security issues.
\subsection{Preliminaries of Multi-tier Computing}
Multi-tier computing is a tiered computing architecture composed of multiple tiers, each with specific functionalities and responsibilities. The goal of this architecture design is to achieve distributed computing, resource sharing, and task offloading to improve computing efficiency, flexibility, and scalability. As shown in Fig. \ref{666}, multi-tier computing typically consists of the following tiers:
\par $\bullet$ Edge tier: The edge tier is located at the edge of the computing system, close to the end users. The edge tier is responsible for receiving, processing, and storing data from end devices. It has lower latency requirements and can perform real-time data processing and decision-making, reducing reliance on the cloud tier.
\par $\bullet$ Cloud tier: The cloud tier is the core tier of the multi-tier computing system, possessing large-scale computing and storage resources. The cloud tier receives data from the edge tier and performs more complex computations and analysis. It provides high-performance computing capability and large storage capacity, capable of handling large-scale datasets and complex computing tasks.
\par $\bullet$ Middle tier: The middle tier is located between the edge tier and the cloud tier, serving as a connector and coordinator. It can preprocess, filter, and aggregate data to reduce data transmission volume and computing load. The middle tier can also provide auxiliary functions such as data transformation, data security, and load balancing.
\par The advantage of the multi-tier computing architecture lies in distributing computing tasks across different tiers, enabling the system to better adapt to different application scenarios and requirements. The edge tier can quickly respond to requests from end devices, reducing latency and network congestion. The cloud tier provides powerful computing and storage capabilities to support complex computing tasks and large-scale data processing \cite{mortazavi2017cloudpath}, \cite{wang2022noma}. In Table \ref{tab:2}, we provide enabling techniques that can be used for digital twins in multi-tier computing.

\subsection{Computation in Multi-Tier Computing}
The idea behind multi-tier computing architecture is to enable fast and efficient task computing by flexibly developing computing capability across the cloud to things. Multi-tier computing enables distribution of computation load across different tiers around the user to fully utilize the computing resources available along the continuum. Thereby, it extends traditional cloud-centric computing architecture to the distributed edge of the network. Because of the multi-tier computing, components with different delay or energy consumption requirements in DT application components can run at different levels. To meet different delay requirements, cloud computing with huge resources is needed to support time-tolerant tasks, and distributed edge computing with limited resources and closer to the user is needed to support simple time-sensitive tasks. 
\par Through heterogeneous computing resources and collaborative computing architecture, multi-tier computing systems can effectively support a full range of services with different requirements. Based on this, multi-tier computing provides the advantage of low-latency task offloading as well as efficient resource utilization, as it allows tasks to be processed through idle resources at the edge of the network near the end users.  Therefore, for network systems with rich computing resources, multi-tier computing can achieve seamless integration of edge systems and cloud systems, which goes beyond the edge server or cloud server as a separate computing platform. For example, the integration of multi-tier computing and non-orthogonal multiple access (NOMA) is used to realize task offloading in the DT. In \cite{wang2020online}, a NOMA based fog computing framework for the industrial internet of things system is proposed, in which multiple task nodes unload their tasks to multiple computing nodes through the NOMA strategy for task computing, minimizing the energy consumption and delay in the total cost constrained by given communication and task computing. 
\begin{table*}[htbp]
	\centering
	\caption{Multi-tier Computing Enabling Technology for Digital Twin}
	\label{tab:2}  
	\begin{tabular}{m{3cm} m{9cm} m{3cm}}
		\hline\hline\noalign{\smallskip}	
		Enabling Technology & Function & Reference  \\
		\noalign{\smallskip}\hline\noalign{\smallskip}
		Edge Computing &  Digital Twin can obtain sensor data in real-time through edge computing, reducing latency and enhancing responsiveness. & \cite{9996947}, \cite{9854866}, \cite{pushpa2020using} \\
        \hline 
		Cloud Computing & Cloud computing technology provides high-performance computing and storage resources, offering powerful computing and analysis capabilities for digital twin. & \cite{wu2021digital}, \cite{tao2017digital},  \cite{dang2021cloud}, \cite{wang2021research} \\
        \hline 
		Artificial Intelligence and Machine Learning & AI and ML technologies can extract patterns and trends from a large amount of real-time data, predict the behavior and performance of physical entities, and achieve intelligent monitoring, optimization, and prediction of digital twin. & \cite{groshev2021toward}, \cite{zhang2022artificial}, \cite{kharchenko2020combination}, \cite{lv2022artificial}, \cite{rathore2021role}, \cite{chakraborty2021machine} \\
        \hline 
		Big Data Analytics&  Digital twin involves a large amount of real-time data. Big data analytics technology enables in-depth analysis and mining of this data, extracting valuable insights and information. & \cite{erikstad2017merging}, \cite{li2022big}\\
        \hline 
		Virtual Reality and Augmented Reality & Virtual reality and augmented reality technologies provide interactive visualization interfaces for Digital Twin, allowing users to intuitively understand the status and performance of physical entities and perform real-time interactions. & \cite{schroeder2016visualising}, \cite{rabah2018towards}, \cite{kunz2022potential}, \cite{yin2023state}  \\
        \hline 		
        Internet of Things &  IoT technology connects sensors and devices, establishing a data exchange channel between physical entities and digital twin, facilitating real-time data collection, transmission, and analysis. & \cite{al2020digital}, \cite{minerva2020digital}, \cite{souza2019digital} \\
        \hline 
		Data Security and Privacy Protection & Data security and privacy protection technologies can protect private information in the digital twin, reduce the risk of privacy leakage, and ensure the security and reliability of the digital twin. &\cite{alcaraz2022digital}, \cite{karaarslan2021digital}  \\
		\noalign{\smallskip}\hline
	\end{tabular}
\end{table*}
\par During the task offloading process of multi-tier computing-based DT, computation and communication resource allocations optimization requires solving a non-convex problem. For example, in the massive multiple-input multiple-output (MIMO)-aided multi-tier computing system, task nodes (TN) unload their tasks to nearby massive MIMO-aided relay nodes (MRN), and the cache resources of these nodes are not fully utilized. Similarly, caching commonly used services can reduce task execution latency. By utilizing the complex relationship between task execution latency and service cache, as well as optimal power allocation, the intertwined service cache and computing resource allocation can be jointly optimized. However, due to the non convex characteristics of the generated optimization problem, finding the optimal solution becomes a challenge \cite{9676649}. Nowadays, alternating optimization technology is a commonly used and effective method to solve optimization problems involving coupling variables. The idea is to decompose a complex non-convex problem into multiple sub-problems and alternately solve these sub-problems until convergence conditions are reached. Each sub-problem usually only involves a portion of the optimization variables, making the solution of the problem more simplified. Through alternating optimization techniques, the performance of multi-tier computing system can be gradually improved in each iteration and gradually approach the global optimal solution. Although each iteration may only obtain a local optimal solution, the cumulative effect of multiple iterations can achieve better overall performance. In \cite{9676649}, an efficient alternating optimization algorithm is proposed to solve the non convex power allocation optimization problem, transforming it into a linear optimization problem with given task offloading and service caching results. In addition, distributed optimization algorithms for non convex optimization have also been widely used in recent years. For example, distributed gradient descent, distributed sub-gradient descent, distributed stochastic optimization, and distributed augmented Lagrangian methods. These algorithms are widely used in distributed systems to deal with non-convex optimization problems, and can achieve the goal of global optimization through information exchange and iterative updating. They can solve non convex optimization problems in multi node and distributed computing environments, and have demonstrated good performance and feasibility in practical applications \cite{8737489}, \cite{doan2020fast}, \cite{amiri2020machine}, \cite{lv2020differentially}.
\par Due to increasingly complex wireless networks, the number of edge nodes and configurable parameters have significantly increased. Multi-tier computing can optimize computing resource allocation through the AI technology, and it is applied in multiple protocol tiers (for example, physical tier resource allocation, data link tier resource allocation and service control) \cite{yang2019multi}, \cite{mao2017survey}.
Edge servers can provide low latency AI services to end users, but these services cannot be directly implemented on devices. To enable low latency task processing, distributed AI, such as federated learning, can be executed on multiple layers of computing nodes in a network. Due to the data being processed by edge servers near smart devices, there is no need to transmit large amounts of data to remote servers \cite{liu2022bringing}, \cite{shi2020communication}. Therefore, the use of edge AI in DT systems can not only save the bandwidth consumption of data upload, but also greatly reduce the execution delay of computational tasks. Edge AI will play a crucial role in reducing the delay of computing tasks by intelligently implementing task offloading and resource allocation decisions.

\subsection{Communication in Multi-Tier Computing}
The multi-tier computing system distributes communication capabilities anywhere from the cloud to the end devices to fully utilize the available communication resources along the continuum. For example, in large-scale multiple-input multiple-output (MIMO) assisted multi-tier computing system, as the antenna size gets larger and the channel becomes more deterministic, the allocation of communication resources depends largely on the distance between the user and the base station. This means that communication resources for DT models do not need to be updated and allocated frequently, thus greatly saving signal transmission costs. This will improve the spectrum and energy efficiency of communications and support an increase in the number of access entities. However, for the real-time requirements of DT, reducing the communication resource overhead through multi-tier computing requires maintaining a certain data rate. Taking the DT system of the IoV as an example, each RSU sends and receives up to Gbits of data per second \cite{xu2020service}. At this point, edge nodes can increase data communication rates by utilizing next-generation communication technologies such as massive MIMO, intelligent reflecting surface, or space-air-ground integrated network \cite{mao2021ai}.
\par DT can also cache and prefetch relevant data on the target device or edge node to reduce transmission delay. At the same time, edge nodes adopt intelligent routing algorithms and transmission scheduling strategies to select the optimal transmission path and time, and realizes communication with different network infrastructures through multi-hop and multi-connection mechanism to reduce transmission delay. 
For example, in the wireless sensor network (WSN) used for DT sensing environment and data collection, the energy required for communication is an important challenge, and rapid energy consumption and network inequity can cause the loss of large amounts of data, resulting in degraded node performance and increased packet distribution latency. Therefore, it is urgent to check the energy usage of nodes in order to make effective routing decisions and improve the overall network performance by applying intelligent machine learning technology. In \cite{thangaramya2019energy}, a new neural fuzzy rule-based cluster formation and routing protocol for efficient routing in IoT based WSN is proposed to provide better network performance in indicators such as energy utilization, packet distribution rate, latency, and network lifespan. The current DT data communication protocol is based on IoT devices connected to the cloud. However, with the introduction of multi-tier computing in the IoT, all processing occurs at the edge, and data is analyzed and filtered before being sent to the cloud. Remote proxies are implemented in edge nodes, and primary proxies are implemented in the cloud. Therefore, it is necessary to re-examine existing IoT protocols. For example, the most commonly used IoT protocol is the message queuing telemetry transmission protocol (MQTT) based on the publish/subscribe model, which allows devices to communicate with agents residing in the cloud. However, it creates more communication overhead for the client. In \cite{veeramanikandan2019publish}, a software-defined multi-tier edge computing model was implemented by modifying the existing MQTT protocol. This enhanced the relevant functions of remote agents to perform various real-time analyses during the critical events at the fog/edge level, enabling the proposed system to work in parallel in a heterogeneous environment with very low transmission delay, low bandwidth requirements and less network congestion.

\subsection{Data Storage in Multi-Tier Computing}
The DT storage data mainly consists of two parts, data from physical space and data from virtual space. Data from physical space is collected and stored by sensors from the edge, but due to the limited storage resources of edge nodes, they can leverage a large number of cloud-like nodes in multi-tier computing to achieve collaborative storage of data. Since multi-tier computing systems can integrate a large number of densely distributed devices, they have elastic storage capacity and dynamically adjust storage resources according to demand. The cache resources of edge servers are usually accessed by smart devices nearby. As a result, the cache resources of the edge nodes may be consolidated into special capacity areas. Multiple interconnected edge infrastructures co-existing in space and time can then pool storage resources for sharing by DT and end users. 
\par In \cite{xing2018distributed}, a distributed multi-level storage architecture with a multiple-factors least frequently used (mLFU) algorithm is proposed. In this architecture, the storage level is composed of intelligent devices at the edge. Therefore, when the storage capability of a node is insufficient, mLFU is used to delete some data from the current node and upload the data to a higher storage level. In order to reduce the impact of data loss caused by unstable edge networks, an important factor was introduced in mLFU, which is to first upload highly important data to the upper tier. Mobile edge caching is an emerging distributed storage paradigm that enables caching at the edge of a network, whether it is at edge base stations or on mobile devices \cite{sabella2016mobile}. Therefore, end users can obtain cached data from adjacent edge servers or adjacent devices instead of remote cloud servers, greatly reducing the burden of peak hour traffic, business latency, and backhaul links to support various applications and services in DT. However, frequent movement of devices with cached data may lead to data loss. Therefore, caching massive amounts of data with high data tolerance and low cost has become an important but highly challenging issue. In \cite{8786259}, a multi-tier mobile edge computing (MEC) network framework is proposed based on repairable fountain code for data caching, repair, and download. Cloud, BS, edge BS, and mobile devices collaborate to store and deliver content to users. This framework enables efficient data caching in multi-tier heterogeneous MEC networks.
\par However, the environment of the physical entity mapped by DT is changeable. The different situations of the request generation area and the change of the overall number of requests lead to the instability of the MEC environment. Due to sudden requests and changing MEC environments, a sudden increase in computation load and a significant load imbalance between MEC servers can also lead to congestion in edge network links in some areas. In addition to compressing data and caching data, the multi-tier computing systems can utilize deep reinforcement learning for data storage and computing resource allocation. In \cite{8657791}, a deep reinforcement learning based intelligent resource allocation scheme is proposed, which can adaptively allocate computing and storage resources, reduce average service time, and balance resource usage in different MEC environments.

\subsection{Privacy and Security in Multi-Tier Computing}
Multi-tier computing can help solve privacy and security issues in DT. Specifically, multi-tier computing can effectively protect privacy information in DT systems by separating and isolating data processing and computing tasks at different levels, as well as adopting security mechanisms such as encryption and access control.
\subsubsection{Data isolation and encryption}
Multi-tier computing can reduce the risk of sensitive data exposure by separating data storage and processing into different computing nodes or systems. At the same time, encryption technology is used to encrypt data to ensure its security during transmission and storage. For example, in \cite{shen2021secure}, a blockchain based DT big data (BDTD) security sharing framework is considered. Cloud storage was integrated into the framework, BDTD was encrypted and stored in the cloud, and the hashes of BDTD and transaction records were stored in the blockchain. The high-speed security sharing of BDTD was realized through a multi-level computing system. In \cite{9830718}, a DT privacy protection communication scheme is designed based on cloud computing network tier and blockchain. Through mutual authentication and key protocol scheme between each node, data owners and cloud storage, data owners and data users can ensure their secure communication in public channels. The protocol can guarantee various security features, including anonymity, untraceability, and perfect forward secrecy.

\subsubsection{Access control and authentication}
Multi-tier computing  can adopt access control policies and authentication mechanisms to restrict access to sensitive data and computing tasks in DT system. Only authorized users or nodes can access and operate related data and computing resources \cite{ruj2012privacy}. In \cite{shiny2022decentralized}, a decentralized access control scheme is aimed at securely storing data on the cloud through multiple key distribution centers and multi-tier user authentication. Combine steganography in cryptography to enrich the security of data to be stored in the cloud. In the proposed organization, regardless of the user's identity, the cloud will confirm the authenticity of the sequence before storing the data. This scheme also has additional elements of access control, allowing only legitimate clients to decrypt retained data and view equivalent data. It is also based on multi-tier authentication to extend user authentication. This decentralized access control design idea in multi-tier computing is applicable to DT. It provides a secure link for a large number of data requests from different physical entity and the feedback of the prediction results of DT.
\subsubsection{Security protocols and communication}
Multi-tier computing can adopt security protocols and communication mechanisms, such as secure transport layer protocol and virtual private network, to protect the data transmission and communication process in the DT system and prevent data from being stolen or tampered with \cite{thakur2023effective}. The physical entity on the edge node and the DT on the cloud can evaluate and select different transport tier protocols for communication according to specific application scenarios and needs. For example, in the industrial control system, we can consider using the open platform communication unified architecture (OPCUA) \cite{9211998}. It has the advantages of platform independence, scalability, support for comprehensive information modeling, communication security, etc. It is a reliable interoperability standard for information exchange in many industries (such as industrial automation). OPCUA consists of 14 specifications and some supporting specifications, defining the interfaces between clients and servers, as well as between servers, including real-time data access, alarm monitoring, etc. The communication and computing capabilities of participating DT devices often differ greatly, and the transmitted data also exhibits heterogeneity. Transmission control protocol can provide reliable data transmission and end-to-end connectivity \cite{8807890}. It can communicate data between different network devices and has functions such as flow control and congestion control. Device network environments located at edge nodes typically have low bandwidth and instability, and lightweight MQTT can be used to ensure the secure transmission of DT data. By subscribing and publishing messages, real-time data transmission and communication in the DT system can be achieved, but it has the characteristics of high latency and high sampling rate, which will affect the prediction results of the DT model \cite{naik2017choice}.
\subsubsection{Security audit and monitoring}
Multi-tier computing can implement security auditing and monitoring mechanisms, monitor and record data access, computing operations, and security events in the system, and timely detect and respond to security threats \cite{bleikertz2010security}. The DT system should record audit logs of all critical operations and events. These logs include user access, data modifications, system configuration changes, etc. The recording of audit logs can track and trace the responsible entities that change simulation parameter settings or status data, and help rebuild the process chain, and finally detect and locate faulty nodes in the system. We can utilize the combination of cloud computing and blockchain for audit logging. The immutability of blockchain provides a guarantee for the security of log content. For example, in \cite{ali2022bcals}, a blockchain based log management system is proposed. The proposed system ensures the security of audit logs, ultimately strengthening users' trust in the computing environment and making it unbreakable even by administrators. The real-time monitoring mechanism of multi-tier computing can also be used to monitor the operational status and security events of DT systems in real-time, enabling timely detection and response to potential security threats. For example, using tools such as the security information and event management system can help achieve real-time monitoring and analysis \cite{khan2019survey}. At the same time, DT systems need to conduct regular vulnerability scanning and security assessment to promptly fix security vulnerabilities in the system and applications. The process of vulnerability management and repair should be recorded and tracked to ensure the continuous security of the system.
\par By comprehensively utilizing these security measures, multi-tier computing can provide protection and control of privacy information in DT systems, reduce the risk of privacy leakage, and ensure the security and credibility of DT systems.

\section{Research Directions and Open Problems}
Multi-tier computing for DT is an emerging technology with many open research problems that need to be addressed by the research community. This section discusses and identifies several open issues in the development of DT-empowered by multi-tier computing systems.

\subsection{System Complexity}
DT and multi-tier computing are both complex technologies and systems, and combining them further increases the complexity. A comprehensive system design and analysis is required, taking into account the characteristics and requirements of DT, and combining the multi-tier computing architecture and algorithms to ensure the reliability, scalability, and manageability of the system.

\subsection{Privacy Protection and Security Verification}
DT involves sensitive physical and virtual data, while multi-tier computing involves data transmission and sharing between different tiers and nodes, which increases the risk of security and privacy protection. In multi-tier computing systems, different parts of computing nodes may have different privacy protection capabilities, so compute nodes should be selected with the best scheme to protect data privacy. However, distributed systems in multi-tier computing are generally more vulnerable to attackers than cloud systems, and the security structure of compute nodes in multi-tier computing systems is usually not as strong as in the cloud. As a result, these nodes may not have enough resources to protect their privacy as the cloud does. Additionally, due to limited resources, edge computing nodes may not be capable of detecting threats. In short, data privacy protection of multi-tier computing nodes will be the focus of research on multi-tier computing systems in the future. At the same time, we also need to consider developing security detection standards for DT models and designing verification kits for the security of DT system, such as probabilistic model checking \cite{shaikh2023probabilistic}, \cite{kwiatkowska2009prism}.

\subsection{Data Consistency}
DT involves a large amount of data exchange and sharing, while multi-tier computing involves distributing computing tasks at different tiers and nodes. In this situation, ensuring data consistency between tiers levels and nodes in the DT system becomes more complex \cite{zhang2022digital}. If data changes at a certain level or node, other tiers and nodes also need to update the data accordingly to ensure the overall consistency of the system. Appropriate mechanisms and algorithms need to be designed to ensure data consistency, such as data synchronization and update strategies. But for the update strategy, it is necessary to consider specific requirements and system characteristics, as well as minimize the cost and delay of data transmission. Designing a universal update strategy algorithm remains a challenge.

\subsection{Multi-Tier Task Allocation}
The construction of a DT system relies on the utilization of technologies such as big data, AI, and IoT. It is necessary to be mindful of efficient consumption of hardware resources, communication resources, computing resources, storage resources, etc. The multi-tier computing system provides additional computing capability for DT at the edge and middle tiers of the network. Thus, managing task allocation across multiple tiers becomes a core issue; that is, how to determine whether a virtual task should be executed in a twin end-user device, a twin fog/edge system, or a twin cloud. In order to achieve efficient task offloading, computing tasks need to be scheduled to computing nodes with different computing capabilities according to different computing models, transmission bandwidth and channel quality. Therefore, the heterogeneity of twin nodes has become a key factor in the design of multi-tier computing empowered DT 
 \cite{wang2019hetmec}. Handling different task computations and various communication resources to manage task offloading has become a major problem for DT.

\section{Conclusion}
In this paper, we introduced DT and its application scenarios, and analyzed its challenges. We proposed a multi-tier computing framework to address limited computing resources, high communication and computation latency, as well as security issues in DT. The paper provides valuable reference and guidance for further promoting the development of the theories, algorithms and systems of multi-tier computing enabled DT.


\ifCLASSOPTIONcaptionsoff
  \newpage
\fi



%
\bibliographystyle{IEEEtran} 
\bibliography{IEEEabrv,ref} 
%




\end{document}